\documentclass[aps,prl,
 reprint,
 superscriptaddress,
 longbibliography,
 amsfonts,amssyb,amsmath,preprintnumbers,floatfix,showpacs,
 letterpapaer,byrevtex,citeauthorscript]{revtex4-1}

\usepackage[T1]{fontenc}	
\usepackage[latin1]{inputenc}	
\usepackage[english]{babel}	
\usepackage{mathptmx}		
\usepackage{bm}			
\usepackage{microtype}		
\usepackage{mathtools}		
\usepackage{graphicx}		
\usepackage{xcolor}		
\usepackage{booktabs}		
\usepackage{comment}		

\RequirePackage{xspace}
\newcommand*{\eg}{\emph{e.g.}\xspace}
\newcommand*{\viz}{\emph{viz.}\xspace}

\newcommand*{\pdd}[2]{\frac{\partial{#1}}{\partial{#2}}}

\newcommand*{\dalembert}{\square}
\newcommand*{\eps}{\varepsilon}
\newcommand*{\e}{\text{e}}

\providecommand{\mbf}{\mathbf}

\newcommand*{\unit}[1]{\hat{\mbf{#1}}}

\providecommand{\hm}{\bm}
\newcommand*{\grad}{\hm{\nabla}}
\newcommand*{\bdot}{\hm\cdot}

\begin{document}

\title{General relativistic massive vector field effects in Gamma-Ray Burst
production}

\author{Fabrizio Tamburini}
\email{fabrizio.tamburini@gmail.com}
\affiliation{Scientist in Residence at ZKM -- Centre for Art and Technologies for Media,
Lorentzstra{\ss}e~19, D-76135 Karlsruhe, Germany}

\author{Mariafelicia De Laurentis}
\email{laurentis@th.physik.uni-frankfurt.de}
\affiliation{Institute for Theoretical Physics,
 Goethe University, Max-von-Laue-Str.~1, D-60438 Frankfurt, Germany}
\affiliation{Frankfurt Institute for Advanced Studies,
 Goethe University, Ruth-Moufang-Str.~1, D-60438 Frankfurt, Germany}

\author{Lorenzo Amati}
\email{amati@iasfbo.inaf.it},
\affiliation{Istituto Nazionale di Astrofisica --- IASF Bologna,
 via P.~Gobetti~101, I-40129 Bologna, Italy}
\affiliation{International Center for Relativistic Astrophysics,
 Piazzale della Repubblica~2, I-65122, Pescara, Italy}

\author{Bo Thid\'e}
\email{bt@irfu.se}
\affiliation{Swedish Institute of Space Physics,
 {\AA}ngstr\"{o}m Laboratory, P.\,O.~Box~537, SE-75121, Sweden} 
\affiliation{Acreo Swedish ICT AB,
 P.\,O.\ Box 1070, SE-16425 Kista, Sweden}

\begin{abstract} 

To explain the extremely high energy release, $>10^{53}$~erg, suggested by
the observations of some Gamma-Ray Bursts (GRBs), we propose a new energy
extraction mechanism from the rotational energy of a Kerr-Newman black
hole (BH) by a massive photon field.  Numerical studies show that this
mechanism is stable with respect to the black hole rotation parameter,
$a$, with a clear dependence on the BH mass, $M$, and charge, $Q$, and
can extract energies up to $10^{54}$~erg. The controversial ``energy
crisis'' problem of GRBs that does not show evidence for collimated
emission may benefit from this energy extraction mechanism.  With these
results we set a lower bound on the coupling between electromagnetic
and gravitational fields.

\end {abstract}


\maketitle

The exact Gamma-Ray Burst (GRB) energy production mechanism 
is still a matter of debate.  One possible explanation is the
energy release during the formation of a rotating black hole
(BH) surrounded by the matter of a rapidly collapsing massive
star and, to account for the high energy observed, one has
to find very efficient energy extraction mechanisms from the BH~%
\cite{vanPutten&al:PRD:2011,%
vanPutten&al:AA:2011,%
Fraser-McKelvie&al:MNRASL:2014}.
Other possible mechanisms can be high-energy phenomena of energy
pulse from electron-positron pair production and the photon
plasma fluid created by vacuum polarization in the dyadosphere~%
\cite{Preparata&al:AA:1998,%
Ruffini:AASS:1999},
a region around a charged BH that extends from the exterior event horizon
$r_+$ to a given radius $r_\text{ds}$ that depends on the mass and charge
of the BH~%
\cite{Damour&Ruffini:PRL:1975,%
Bini&al:PLA:2007,%
Bini&al:PRD:2007,%
Ruffini&Xue:AIP:2008}.

Vacuum polarization effects require strong magnetic fields that mimic the
effects of a charged and rotating BH described by the Kerr-Newman solution~%
\cite{Chandrasekhar:Book:1992}.
However, because typical astrophysical systems show a strong tendency
to eliminate any net electric charge, a charged BH is not a realistic
physical solution and cannot be generated by the gravitational collapse
of a core larger than the neutron star critical mass endowed with an
electromagnetic field.  A reasonable physical solution is to consider
energy extraction mechanisms where a BH that ``acquires'' a temporary
fictitious net charge because of vacuum polarization effects and/or
generates an electrostatic field that extends towards the neighborhoods of
the event horizon through a selective capture of charged particles. This
temporary fictitious charge is expected to dissipate on time scales
$\tau<10^7$~s, long enough to set up the GRB energy extraction
mechanisms occurring on much shorter time scales during the collapse
of the star and thus justify the use of Kerr-Newman based spacetimes~%
\cite{Preparata&al:AA:1998,Ruffini:AASS:1999}.

As described in Ref.~\onlinecite{Dolgov&Pellicia:PLB:2007}, photons can
acquire mass due to gravitational and electromagnetic field coupling.
Exploiting this possibility, we here describe a new mechanism of energy
extraction from black-hole rotational energy made possible by vacuum
polarization effects occurring in the dyadosphere around a Kerr-Newman BH.

The main motivation for this work arises from observations of GRBs that
do not show evidence of collimated emission~%
\cite{Panaitescu&al:MNRAS:2006,%
Campana&al:AA:2007,%
Liang&al:APJ:2008,%
Kann&al:APJ:2010,%
Kann&al:APJ:2011}.
As shown, \eg, by \citeauthor{Amati&DellaValle:IJMPD:2013}~%
\cite{Amati&DellaValle:IJMPD:2013},
GRBs have a distribution of released energy, in terms of equivalent
isotropic radiation peaking at about $E_\text{iso}\sim10^{53}$~erg or
higher, \eg, GRB~130427A; see~%
\cite{DePasquale&al:ARXIV:2016}.
In some cases they seem to be able to have up to
$E_\text{iso}\sim10^{54}$~erg, corresponding to emitting one Sun rest
mass energy in $1$--$2$ minutes.

The collimated emission scenario often called for to solve this
``\emph{energy crisis}'' implies the detection in the light curves of
the afterglow of a sudden ``\emph{achromatic'}' change in the slope
(break) at the time when the relativistic beaming angle becomes larger
than the physical jet opening angle as a consequence of the slowing down
of the ejected shells~%
\cite{Sari&al:APJL:1999}.
Based on this method, beaming angles between $4$ and $9$ degrees~
can be estimated for a number of GRBs~%
\cite{Frail&al:APJL:2001,%
Ghirlanda&al:MNRAS:2007,%
Ghirlanda&al:AA:2007,%
Liang&al:APJ:2008},
decreasing their energy budget to below $\sim5\times10^{52}$~erg.
This is comparable to the kinetic energy of so called hypernovae,
a sub-class of SNe-Ibc often associated with GRBs~%
\cite{DellaValle:IJMPD:2011}.

\begin{figure}
\thicklines
\includegraphics[width=\columnwidth]{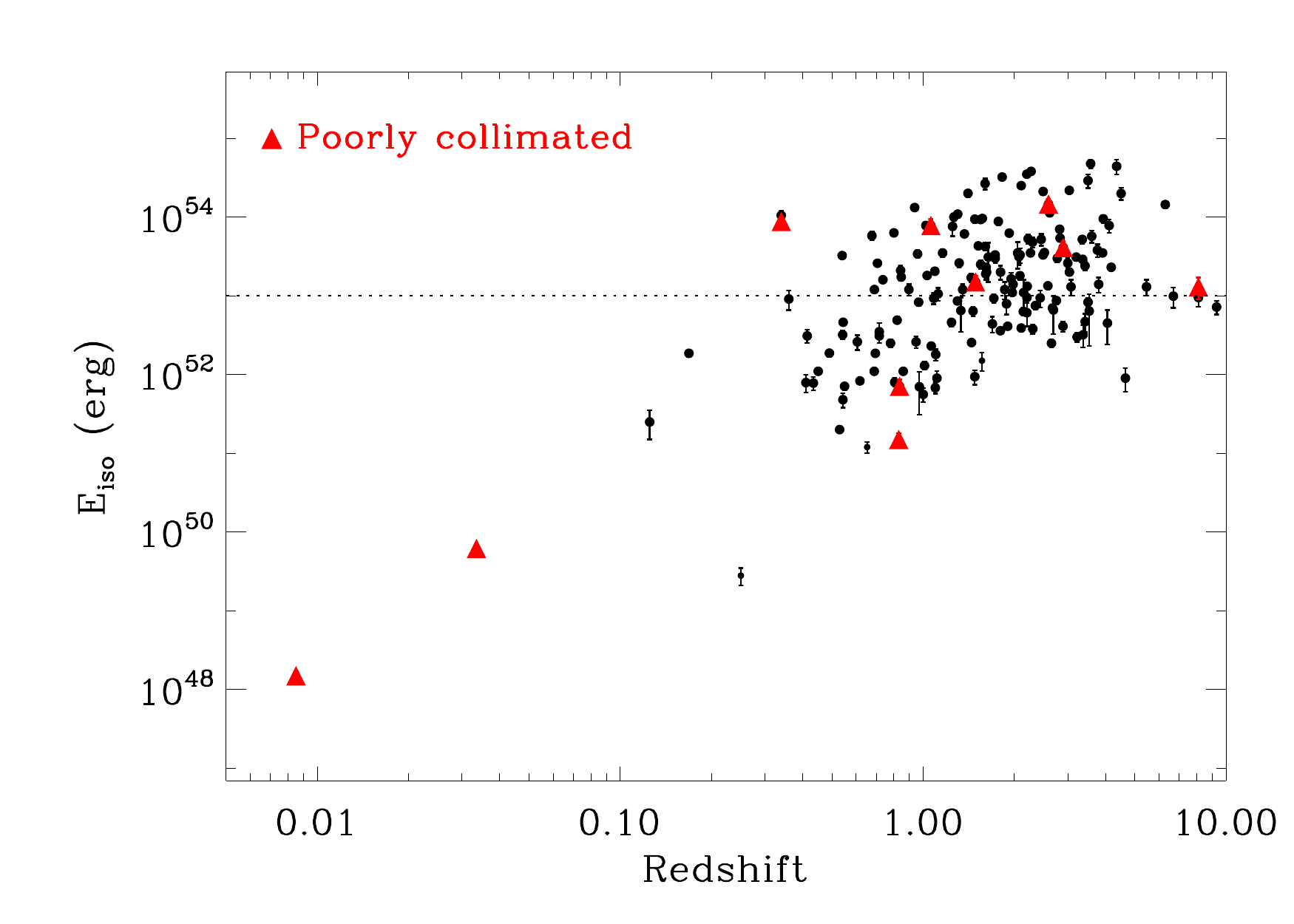}
\caption{Isotropic radiated energy distribution of GRBs, $E_\text{iso}$,
 w.r.t.~their redshift (see Ref.~\onlinecite{Amati&DellaValle:IJMPD:2013}
 and references therein). Red triangles indicate those GRBs that exhibit
 almost spherical emission (sub-energetic, \eg, GRB~980425 and GRB~060218~%
 \cite{Guetta&DellaValle:APJL:2007})
 or those whose optical follow-up provided evidence of no, or a late,
 characteristic break (see the compilation by~%
 \citeauthor{Chandra&Frail:APJ:2012} \cite{Chandra&Frail:APJ:2012}%
 ) and thus are supposed to have a low degree of collimation (see text). In
 this case $E_\text{iso}$ can give the correct order of magnitude of the
 radiated energy.}
\label{fig:isotropic}
\end{figure}

However, in some cases this change in the slope is not observed~%
\cite{Campana&al:AA:2007,%
Liang&al:APJ:2008,%
Kann&al:APJ:2010,%
Kann&al:APJ:2011},
suggesting that some GRBs could be characterized by poorly collimated
emission. A quasi-isotropic emission is indeed often observed in
low-luminosity GRBs~%
\cite{Guetta&DellaValle:APJL:2007},
but in this case the energy crisis is avoided by the low amount of
energy associated with $E_\text{iso}\sim10^{50}$~erg, as shown in
Fig.~\ref{fig:isotropic}.  However the existence of high-luminosity GRBs~%
\cite{Guetta&DellaValle:APJL:2007,%
Guetta&Piran:JCAP:2007}
that do not exhibit an ``\emph{achromatic}'' change in the slope of the
afterglow lightcurves, poses the question to pinpoint an alternative
mechanism of energy production in GRBs other than the ones normally
adopted, such as ``failed supernovae''~%
\cite{Woosley:APJ:1993,%
Woosley&Bloom:ARAA:2006}
and magnetars~%
\cite{Lyutikov&Usov:APJL:2000,%
Zhang&al:APJ:2004,%
Metzger&al:MNRAS:2010}
and able to justify up to $\sim10^{54}$~erg.

We find that the energy extraction from a Kerr-Newman BH by a massive
photon (vector) field can be as large as $\sim10^{54}$~erg, which
is indeed the upper limit of the energy budget observed in GRB events~%
\cite{Amati&al:AA:2007,%
Capozziello&Lambiase:PLB:2015}.

The presence of an electrostatic charge on the BH unavoidably decreases
the efficiency of the BH energy extraction process if charged particles
are accreting~%
\cite{Bhat&al:JAA:1985}.
For this reason, the charged particles that form the plasma around the
BH cannot provide a fully efficient energy extraction in the Penrose
process that would generate the energy needed to form a GRB.  One way of
increasing the energy emission is to include also the contribution of
additional resources such as accreting massive neutral vector fields.
Massive Proca-Maxwell photons or massive vector field non-minimally
coupled to gravitation due to dark energy~%
\cite{Bohmer&Harko:EPJC:2007}
represent a possibility.

Because of the scenario involved in regions surrounding the BH as
described by the dyadosphere model~%
\cite{Preparata&al:AA:1998,%
Ruffini:AASS:1999},
where the vacuum is polarized and processes of pair production,
annihilation and oscillation lead to the formation of a high-energetic
and dense plasma of electrons, positrons and photons. Because of number
equipartition, electron-positron pairs number densities are asymptotically
comparable to that of photons that obey Proca-Maxwell vector fields.
A massive photon Proca-Maxwell field has no electric charge and non-zero
rest mass under the particular conditions described below. Being neutral
particles, their extraction energy efficiency from the BH is not affected
the by temporary Kerr-Newman electrostatic charge. The electromagnetic
(EM) potential enters actively into the energy budget of the Penrose
process, and the finite rest mass of the photon plays a crucial r\^ole,
changing the spacetime metrics and increasing the total energy extraction.

The generation of a massive photon field requires a space-time
symmetry breaking of Maxwell's equations and a coupling between the
electromagnetic and gravitational field to modify spacetime curvature
and the energy extraction.  This can happen in this scenario, the
core of the collapsing star is thought to be surrounded by a plasma
during the BH formation, photons acquire a virtual mass through the
Anderson-Higgs mechanism~%
\cite{Schwinger:PR:1962,%
Anderson:PR:1963}
as a hidden gauge invariance of the Proca-Maxwell equations preserving
the Lorentz invariance~%
\cite{Mendonca:Book:2001}.
The Proca-Maxwell Lagrangian density $\mathcal{L}$ describing
a massive electromagnetic (EM) field contains a mass term
$\mu_\gamma^2A_\mu A^\mu/2$ because of the gauge invariance [see
Ref.~\onlinecite[][Eq.~(3)]{Tamburini&al:EPL:2010}]
\begin{equation}\label{L}
\mathcal{L}=
 - \frac{1}{4} F_{\mu\nu}F^{\mu\nu}-j_\mu A^\mu
 + \frac{1}{2} \mu_\gamma ^2 A_\mu A^\mu\,,
\end{equation}
where $\mu_\gamma^{-1}$ is the reduced Compton wavelength associated
with the photon rest mass and $j_\mu=\left(\rho,- \mathbf{j} \right)$
is the $4$-current. $F^{\mu\nu}$ is the electromagnetic field tensor
and $A_\mu$ the $4$-vector potential. From the Lagrangian density,
one obtains the covariant form of the Proca-Maxwell equations
\begin{equation}
\label{eq:Proca-Maxwell}
 \pdd{F_{\mu\nu}}{x_\nu}+\mu_\gamma^2A_\mu=4\pi j_\mu\,,
\end{equation}
that lead to the Proca wave equation for the 4-potential $A_\mu$
\begin{equation}
 (\dalembert -\mu_\gamma ^2)A_\mu= - 4 \pi j_\mu\,,
\end{equation}
with the constraint derived from the massive photon in a medium
($\partial^\mu{A}_\mu=0$).  Proca-Maxwell equations include a mass-like
term for the photon due to light-matter interaction, \viz, a term
encompassing the photon's interaction with the plasma (photon-plasmon
interactions).

Photons and plasmas themselves gravitate, but not their coupling,
unless a non-minimal coupling of the EM vector potential to gravity is
present in the Lagrangian, ${\mathcal{L}_R=\xi{R}A^\nu{A}_\nu}$, where
$R$ is the curvature scalar and $\xi$ a coupling constant. Because of
this, photons carry a finite rest mass, and an additional term appears
in Einstein's equations.  This coupling is thought to occur, \eg, in
the presence of a charge asymmetry~%
\cite{Dolgov&Pellicia:PLB:2007}
as is supposed to happen in regions near the forming BH of a GRB.
An exhaustive theoretical description of this coupling is beyond the
scope of this Letter and would need a deep understanding of a possible
unification between electromagnetism and gravitation.

The stress-energy tensor of Proca-Maxwell equations would not be traceless
like that of Maxwell's equations, but due to symmetry breaking and EM and
gravitational coupling, the total photon ``Proca mass'' would introduce an
additional mass term, being intrinsically gravitating and thus included
as a source in Einstein's equations.  When photons acquire such an
additional mass term, the spacetime curvature is unavoidably modified
and the environment around the BH is endowed with a massive vector
field described by the Proca-Maxwell equations of the EM field. See,
\eg, Refs.~%
\onlinecite{%
Bei&al:IJTP:2004,%
Shi&Liu:IJTP:2005,%
Pani&al:PRL:2012,%
Pani&al:PRD:2012}
for more details.

If the plasma is turbulent, then one has to consider an additional
effect due to the spatial structure of the plasma itself at the plasma
resonance frequencies. In fact, unlike the mathematical structure of
the space-time manifold described by the Lorentz group in which space
is homogeneous and isotropic and time homogeneous, a plasma may exhibit
peculiar spatial/temporal structures that break the space-time symmetry
and generates the conversion of a fraction of the Proca mass into photon
orbital angular momentum (OAM) that acts as a mass reducing  term in
the photon Proca mass~%
\cite{Tamburini&al:EPL:2010,%
Tamburini&Thide:EPL:2011,%
Anderson:PR:1963}.
In this case the OAM term changes the mass in the Proca-Maxwell
equations and the Yukawa potential of the space-time curvature~%
\cite{Capozziello&DeLaurentis:APB:2012,%
Capozziello&DeLaurentis:PHR:2011},
preserving, in any case, the sign of the norm of the metric
tensor, $\|g\|$ as the acquired mass never becomes negative~%
\cite{Tamburini&Thide:EPL:2011}.

Accretion of a massive neutral field in the dyadosphere is much more
efficient than accretion involving charged particles and can supply
or even replace the electromagnetic pulse that is expected to occur to
generate the GRB.  Moreover, the presence of these massive photons modify
the original Kerr-Newman metric by introducing a Yukawa potential that
depends on the electrostatic charge $Q$ of the BH, via the term $\Xi(Q)$
describing the charge asymmetry. For the simplest model of charged BH,
one obtains $\Xi(Q)=Q^2/4 \pi$.

\begin{table}
\caption{Contribution of a massive photon field to the energy produced
by the Penrose extraction process from a rotating charged black hole.}
\label{tab:contribution}
\begin{ruledtabular}
\begin{tabular}{ccccc}
$\mu_{\gamma}$ (g) & $M_{\odot} (g) $ & $a$ & $Q/M$ & Energy \\
$1.78\times10^{-51}$ & $3$ & $0.1$ & $7\times10^{-3}$ & $2.88\times 10^{51}$ \\
$1.78\times10^{-53}$ & $5$ & $0.5$ & $7\times10^{-3}$ & $5.77\times 10^{52}$ \\
$1.78\times10^{-53}$ & $3$ & $0.9$ & $7\times10^{-2}$ & $2.88\times 10^{53}$ \\
\end{tabular}
\end{ruledtabular}
\end{table}

The modified Kerr-Newman metric in natural units where ${e=G=c=1}$,
and in the Boyer-Lindquist coordinates $(t, r, \theta, \varphi)$, with
the presence of the massive photon field also carrying OAM, becomes
\begin{widetext}
\begin{multline}
ds^2 =
 -\Big(
   1-\frac{2Mr+\Xi(Q)\e^{-\mu_{\gamma T}r}}{\rho^2}
   +\mu_{\gamma T} I_{\Xi}
  \Big)\,dt^2 
 +\Big(
   \frac{r^2+a^2}{\rho^2}
   - \frac{2Mr+\Xi(Q)\e^{-\mu_{\gamma T}r}}{\rho^2}
   +\mu_{\gamma T}I_{\Xi}
  \Big)^{-1} dr^2
\\
 +\rho^2 d\theta^2
 +\Big[
   \rho^2\sin^2\theta
   +\Big(
    \frac{2Mr+\Xi(Q)\e^{-\mu_{\gamma T}r}}{\rho^2}
    \mu_\gamma I_{\Xi}
   \Big)
   a^2\sin^2\theta
  \Big]\,d\varphi^2
\\
 +2a\sin^2\theta
  \Big(
   \frac{2Mr + \Xi(Q) \e^{-\mu_{\gamma T}  r}}{\rho^2}
   -\mu_{\gamma T}I_{\Xi}
  \Big)\,d\varphi\,dt
\end{multline}
\end{widetext}
where $\rho^2=r^2+a^2\cos^2\theta$ and
$I_{\Xi}=\int_r^\infty\Xi(Q)(\e^{-\mu_{\gamma T}r}/\rho^2)\,dr$.

To the first order, when $h=1$, the metric shows a dependence on the absolute value
of the OAM acquired by photons \cite{Tamburini&al:EPL:2010}, \viz,
\begin{equation}
\label{eq:firstorder}
 \mu_{\gamma T}\sim 2 \pi \left(P_\mu
 -\ell\,\frac{D_\mu \tilde{n}\sin(q r)}{2P_\mu} \right)\,,
\end{equation}
and the parameters characterizing the Proca mass are
\begin{equation}
 P_\mu =  \mu_G + \sqrt{B_\mu+C_\mu  - D_\mu \cos(q r)}\,,
\end{equation}
where
\begin{subequations}
\begin{align}
\label{eq:Bmu_Cmu_Dmu}
 B_\mu
 &= E\omega^2_{p0}\,
    \frac{1+\eps}{E+\unit{v}\bdot\grad\phi}
\\
 C_\mu
&= \frac{4\pi\delta\dot{v}n_0-4\pi\unit{v}\bdot\dalembert\grad\phi}
        {E+\unit{v}\bdot\grad\phi} 
\\
 D_\mu
&= \frac{4\pi\varphi^\ast\delta\dot{v}}
         {E+\unit{v}\bdot\grad\phi}\,.
\end{align}
\end{subequations}

If the EM field and the plasma density tend to zero, together with $\ell$,
the virtual and actual photon mass, $\mu$ and $\mu_T$ go to zero thus
recovering the Kerr-Newman spacetime.

For a BH with mass in the range of $3$--$5~M_\odot$, electric charge
$Q/M\sim7\times10^{-3}$--$7\times10^{-2}$ and rotation parameter $a\leq1$,
we find that the Yukawa potential is always confined between the radii
$r_1=2.95\times10^{6}$~cm and $r_2=2.81\times10^7$~cm.  The numerical results
indicate that energy produced by the Penrose extraction process of the
massive photon field provides the lacking of energy budget required to
reach the $10^{54}$~erg needed for isotropic emission, having the same
order of magnitude as in the GRB process, $10^{47}$--$10^{54}$~erg,
as reported in Tab.~\ref{tab:contribution}  The lower bounds on photon
mass values obtained by varying $\ell$ as a free integer parameter,
are those derived from massive vector fields, according to Refs.~%
\onlinecite{Pani&al:PRL:2012,%
Pani&al:PRD:2012}
in the range of $\mu_\gamma\sim10^{-53}$--$10^{-51}$~g.

We observe a strong dependence on the BH mass $M$ and charge $Q$, whereas
the BH rotation parameter, $a$, plays a minor r\^ole.  These values
allow us to set the lower bound for the coupling constant between the
EM and gravitational field to $\xi\sim10^{-38}$.

Accretion onto a Kerr-Newman black hole of massive neutral vector fields
can explain the ``energy crisis'' problem of GRBs.  Photons propagating
in a structured plasma and in a strong gravitational field acquire mass
and orbital angular momentum because of the hidden gauge invariance due to
the Anderson-Higgs mechanism in Proca-Maxwell equations when there is an
effective coupling between the gravitational and electromagnetic fields.
The interplay between BH mass, charge, and rotation for the energy
extraction from the neutrally-charged massive photon field gives the
correct order of magnitude of the energy expected for a GRB, showing
a strong dependence on the charge $Q$, with the result of diminishing the
importance of the rotation parameter in the energy extraction from the BH.
This effect, similar to that provided by charged particles falling into
the BH~%
\cite{Bhat&al:JAA:1985},
could flatten the possible differences in the energy distribution due
to differently rotating BHs.

Superradiant instabilities because of the small mass of the BH in
the GRB precursor considered in our calculations do not play a crucial
r\^ole in the too rapid formation of the GRB. The effects induced by the
plasma turbulence on the photon mass, effective at the frequencies where
the plasma is resonant, seem not to play a significant r\^ole in the
energy extraction.  Moreover, this study could also be used in the energy
budget in massive neutrino-antineutrino fireball models if the couples
so generated experience both a Penrose process and viscous dissipation~%
\cite{Shapiro&al:NATO:2005}
or in the accretion of dark matter vector fields.

From our results we can set a lower bound for the coupling constant
between the EM and gravitational field to $\xi\sim10^{-38}$ that might
contribute to electromagnetic emission also in those scenarios where EM
waves are not expected, such as in BH-BH collisions, as hypothesized for the
recent LIGO GW150914 gravitational wave detection and possible Fermi event
coincidence observed after $0.4$~s in the electromagnetic spectrum.  
\cite{Abbott&al_LIGO:ARXIV:2016,%
Abbott&al_LIGO:PRL:2016,Loeb:AJP:2016}.

The effect of this
field coupling might also be accompanied with other non-linear gravitational
wave interactions with plasmas~%
\cite{Brodin&al:PRD:2000}.
No direct transfer of OAM from Kerr metric lensing to photons~%
\cite{Tamburini&al:NPHY:2011} has been considered because of the physical
properties inside the dyadosphere.

The authors thank Massimo Della Valle for the invaluable help on
writing this work.  M.\,D.\,L.\ is supported by the ERC Synergy Grant
``BlackHoleCam'' -- Imaging the Event Horizon of Black Holes (Grant
No.~610058).  B.\,T.\ was financially supported by the Swedish Research
Council (VR) under the contract number~2012-3297.


\begin{thebibliography}{52}%
\makeatletter
\providecommand \@ifxundefined [1]{%
 \@ifx{#1\undefined}
}%
\providecommand \@ifnum [1]{%
 \ifnum #1\expandafter \@firstoftwo
 \else \expandafter \@secondoftwo
 \fi
}%
\providecommand \@ifx [1]{%
 \ifx #1\expandafter \@firstoftwo
 \else \expandafter \@secondoftwo
 \fi
}%
\providecommand \natexlab [1]{#1}%
\providecommand \enquote  [1]{``#1''}%
\providecommand \bibnamefont  [1]{#1}%
\providecommand \bibfnamefont [1]{#1}%
\providecommand \citenamefont [1]{#1}%
\providecommand \href@noop [0]{\@secondoftwo}%
\providecommand \href [0]{\begingroup \@sanitize@url \@href}%
\providecommand \@href[1]{\@@startlink{#1}\@@href}%
\providecommand \@@href[1]{\endgroup#1\@@endlink}%
\providecommand \@sanitize@url [0]{\catcode `\\12\catcode `\$12\catcode
  `\&12\catcode `\#12\catcode `\^12\catcode `\_12\catcode `\%12\relax}%
\providecommand \@@startlink[1]{}%
\providecommand \@@endlink[0]{}%
\providecommand \url  [0]{\begingroup\@sanitize@url \@url }%
\providecommand \@url [1]{\endgroup\@href {#1}{\urlprefix }}%
\providecommand \urlprefix  [0]{URL }%
\providecommand \Eprint [0]{\href }%
\providecommand \doibase [0]{http://dx.doi.org/}%
\providecommand \selectlanguage [0]{\@gobble}%
\providecommand \bibinfo  [0]{\@secondoftwo}%
\providecommand \bibfield  [0]{\@secondoftwo}%
\providecommand \translation [1]{[#1]}%
\providecommand \BibitemOpen [0]{}%
\providecommand \bibitemStop [0]{}%
\providecommand \bibitemNoStop [0]{.\EOS\space}%
\providecommand \EOS [0]{\spacefactor3000\relax}%
\providecommand \BibitemShut  [1]{\csname bibitem#1\endcsname}%
\let\auto@bib@innerbib\@empty
\bibitem [{\citenamefont {{van Putten}}\ \emph
  {et~al.}(2011{\natexlab{a}})\citenamefont {{van Putten}}, \citenamefont
  {Kanda}, \citenamefont {Tagoshi}, \citenamefont {Tatsumi}, \citenamefont
  {Masa-Katsu},\ and\ \citenamefont {{Della Valle}}}]{vanPutten&al:PRD:2011}%
  \BibitemOpen
  \bibfield  {author} {\bibinfo {author} {\bibfnamefont {Maurice H. P.~M.}\
  \bibnamefont {{van Putten}}}, \bibinfo {author} {\bibfnamefont {Nobuyuki}\
  \bibnamefont {Kanda}}, \bibinfo {author} {\bibfnamefont {Hideyuki}\
  \bibnamefont {Tagoshi}}, \bibinfo {author} {\bibfnamefont {Daisuke}\
  \bibnamefont {Tatsumi}}, \bibinfo {author} {\bibfnamefont {Fujimoto}\
  \bibnamefont {Masa-Katsu}},\ and\ \bibinfo {author} {\bibfnamefont
  {Massimo}\ \bibnamefont {{Della Valle}}},\ }\bibfield  {title} {\enquote
  {\bibinfo {title} {Prospects for true calorimetry on {Kerr} black holes in
  core-collapse supernovae and mergers},}\ }\href {\doibase
  10.1103/PhysRevD.83.044046} {\bibfield  {journal} {\bibinfo  {journal}
  {Phys.\ Rev.\ D}\ }\textbf {\bibinfo {volume} {83}},\ \bibinfo {pages}
  {044046} (\bibinfo {year} {2011}{\natexlab{a}})}\BibitemShut {NoStop}%
\bibitem [{\citenamefont {{van Putten}}\ \emph
  {et~al.}(2011{\natexlab{b}})\citenamefont {{van Putten}}, \citenamefont
  {{Della Valle}},\ and\ \citenamefont {Levinson}}]{vanPutten&al:AA:2011}%
  \BibitemOpen
  \bibfield  {author} {\bibinfo {author} {\bibfnamefont {M.~H. P.~M.}\
  \bibnamefont {{van Putten}}}, \bibinfo {author} {\bibfnamefont
  {M.}~\bibnamefont {{Della Valle}}},\ and\ \bibinfo {author} {\bibfnamefont
  {A.}~\bibnamefont {Levinson}},\ }\bibfield  {title} {\enquote {\bibinfo
  {title} {Electromagnetic priors for black hole spindown in searches for
  gravitational waves from supernovae and long {GRBs}},}\ }\href {\doibase
  10.1051/0004-6361/201118080} {\bibfield  {journal} {\bibinfo  {journal}
  {Astron.\ Astrophys.}\ }\textbf {\bibinfo {volume} {535}},\ \bibinfo {pages}
  {5} (\bibinfo {year} {2011}{\natexlab{b}})}\BibitemShut {NoStop}%
\bibitem [{\citenamefont {Fraser-McKelvie}\ \emph {et~al.}(2014)\citenamefont
  {Fraser-McKelvie}, \citenamefont {Brown},\ and\ \citenamefont
  {Pimbblet}}]{Fraser-McKelvie&al:MNRASL:2014}%
  \BibitemOpen
  \bibfield  {author} {\bibinfo {author} {\bibfnamefont {Amelia}\ \bibnamefont
  {Fraser-McKelvie}}, \bibinfo {author} {\bibfnamefont {Michael J.~I.}\
  \bibnamefont {Brown}},\ and\ \bibinfo {author} {\bibfnamefont {Kevin~A.}\
  \bibnamefont {Pimbblet}},\ }\bibfield  {title} {\enquote {\bibinfo {title}
  {The rarity of star formation in brightest cluster galaxies as measured by
  {WISE}},}\ }\href {\doibase 10.1093/mnrasl/slu117} {\bibfield  {journal}
  {\bibinfo  {journal} {Mon.\ Not.\ Roy.\ Astron.\ Soc.\ Lett.}\ }\textbf
  {\bibinfo {volume} {444}},\ \bibinfo {pages} {L63--L67} (\bibinfo {year}
  {2014})}\BibitemShut {NoStop}%
\bibitem [{\citenamefont {Preparata}\ \emph {et~al.}(1998)\citenamefont
  {Preparata}, \citenamefont {Ruffini},\ and\ \citenamefont
  {Xue}}]{Preparata&al:AA:1998}%
  \BibitemOpen
  \bibfield  {author} {\bibinfo {author} {\bibfnamefont {Giuliano}\
  \bibnamefont {Preparata}}, \bibinfo {author} {\bibfnamefont {Remo}\
  \bibnamefont {Ruffini}},\ and\ \bibinfo {author} {\bibfnamefont {She-Sheng}\
  \bibnamefont {Xue}},\ }\bibfield  {title} {\enquote {\bibinfo {title} {The
  dyadosphere of black holes and gamma-ray bursts},}\ }\href@noop {} {\bibfield
   {journal} {\bibinfo  {journal} {Astron.\ Astrophys.}\ }\textbf {\bibinfo
  {volume} {338}},\ \bibinfo {pages} {L87--L90} (\bibinfo {year}
  {1998})}\BibitemShut {NoStop}%
\bibitem [{\citenamefont {Ruffini}(1999)}]{Ruffini:AASS:1999}%
  \BibitemOpen
  \bibfield  {author} {\bibinfo {author} {\bibfnamefont {R.}~\bibnamefont
  {Ruffini}},\ }\bibfield  {title} {\enquote {\bibinfo {title} {The
  {Dyadosphere} of black holes and gamma-ray bursts},}\ }\href {\doibase
  10.1051/aas:1999331} {\bibfield  {journal} {\bibinfo  {journal} {Astron.\
  Astrophys.\ Suppl.\ Ser.}\ }\textbf {\bibinfo {volume} {138}},\ \bibinfo
  {pages} {513--514} (\bibinfo {year} {1999})}\BibitemShut {NoStop}%
\bibitem [{\citenamefont {Damour}\ and\ \citenamefont
  {Ruffini}(1975)}]{Damour&Ruffini:PRL:1975}%
  \BibitemOpen
  \bibfield  {author} {\bibinfo {author} {\bibfnamefont {Thibaut}\ \bibnamefont
  {Damour}}\ and\ \bibinfo {author} {\bibfnamefont {Remo}\ \bibnamefont
  {Ruffini}},\ }\bibfield  {title} {\enquote {\bibinfo {title} {Quantum
  electrodynamical effects in {Kerr-Newmann} geometries},}\ }\href {\doibase
  10.1103/PhysRevLett.35.463} {\bibfield  {journal} {\bibinfo  {journal}
  {Phys.\ Rev.\ Lett.}\ }\textbf {\bibinfo {volume} {35}},\ \bibinfo {pages}
  {463--466} (\bibinfo {year} {1975})}\BibitemShut {NoStop}%
\bibitem [{\citenamefont {Bini}\ \emph
  {et~al.}(2007{\natexlab{a}})\citenamefont {Bini}, \citenamefont {Geralico},\
  and\ \citenamefont {Ruffini}}]{Bini&al:PLA:2007}%
  \BibitemOpen
  \bibfield  {author} {\bibinfo {author} {\bibfnamefont {D.}~\bibnamefont
  {Bini}}, \bibinfo {author} {\bibfnamefont {A.}~\bibnamefont {Geralico}}, \
  and\ \bibinfo {author} {\bibfnamefont {R.}~\bibnamefont {Ruffini}},\
  }\bibfield  {title} {\enquote {\bibinfo {title} {On the equilibrium of a
  charged massive particle in the field of a {Reissner-Nordstr{\"o}m} black
  hole},}\ }\href {\doibase 10.1016/j.physleta.2006.09.028} {\bibfield
  {journal} {\bibinfo  {journal} {Phys.\ Lett.\ A}\ }\textbf {\bibinfo {volume}
  {360}},\ \bibinfo {pages} {515--517} (\bibinfo {year}
  {2007}{\natexlab{a}})}\BibitemShut {NoStop}%
\bibitem [{\citenamefont {Bini}\ \emph
  {et~al.}(2007{\natexlab{b}})\citenamefont {Bini}, \citenamefont {Geralico},\
  and\ \citenamefont {Ruffini}}]{Bini&al:PRD:2007}%
  \BibitemOpen
  \bibfield  {author} {\bibinfo {author} {\bibfnamefont {D.}~\bibnamefont
  {Bini}}, \bibinfo {author} {\bibfnamefont {A.}~\bibnamefont {Geralico}}, \
  and\ \bibinfo {author} {\bibfnamefont {R.}~\bibnamefont {Ruffini}},\
  }\bibfield  {title} {\enquote {\bibinfo {title} {Charged massive particle at
  rest in the field of a {Reissner-Nordstr{\"o}m} black hole},}\ }\href
  {\doibase 10.1103/PhysRevD.75.044012} {\bibfield  {journal} {\bibinfo
  {journal} {Phys.\ Rev.\ D}\ }\textbf {\bibinfo {volume} {75}},\ \bibinfo
  {pages} {044012} (\bibinfo {year} {2007}{\natexlab{b}})}\BibitemShut
  {NoStop}%
\bibitem [{\citenamefont {Ruffini}\ and\ \citenamefont
  {Xue}(2008)}]{Ruffini&Xue:AIP:2008}%
  \BibitemOpen
  \bibfield  {author} {\bibinfo {author} {\bibfnamefont {Remo}\ \bibnamefont
  {Ruffini}}\ and\ \bibinfo {author} {\bibfnamefont {She-Sheng}\ \bibnamefont
  {Xue}},\ }\bibfield  {title} {\enquote {\bibinfo {title} {Dyadosphere formed
  in gravitational collapse},}\ }in\ \href {\doibase 10.1063/1.3012287} {\emph
  {\bibinfo {booktitle} {AIP~Conf.\ Proc.}}},\ Vol.\ \bibinfo {volume} {1059}\
  (\bibinfo  {publisher} {AIP Publishing},\ \bibinfo {address} {Taipei-Hualien,
  Taiwan},\ \bibinfo {year} {2008})\ pp.\ \bibinfo {pages}
  {72--100}\BibitemShut {NoStop}%
\bibitem [{\citenamefont {Chandrasekhar}(1992)}]{Chandrasekhar:Book:1992}%
  \BibitemOpen
  \bibfield  {author} {\bibinfo {author} {\bibfnamefont {S.}~\bibnamefont
  {Chandrasekhar}},\ }\href@noop {} {\emph {\bibinfo {title} {The mathematical
  theory of black holes}}}\ (\bibinfo  {publisher} {Oxford University Press},\
  \bibinfo {address} {New York, NY, USA},\ \bibinfo {year} {1992})\BibitemShut
  {NoStop}%
\bibitem [{\citenamefont {Dolgov}\ and\ \citenamefont
  {Pelliccia}(2007)}]{Dolgov&Pellicia:PLB:2007}%
  \BibitemOpen
  \bibfield  {author} {\bibinfo {author} {\bibfnamefont {Alexander}\
  \bibnamefont {Dolgov}}\ and\ \bibinfo {author} {\bibfnamefont {Diego~N.}\
  \bibnamefont {Pelliccia}},\ }\bibfield  {title} {\enquote {\bibinfo {title}
  {Photon mass and electrogenesis},}\ }\href {\doibase
  10.1016/j.physletb.2007.05.004} {\bibfield  {journal} {\bibinfo  {journal}
  {Phys.\ Lett.\ B}\ }\textbf {\bibinfo {volume} {650}},\ \bibinfo {pages}
  {97--102} (\bibinfo {year} {2007})}\BibitemShut {NoStop}%
\bibitem [{\citenamefont {Panaitescu}\ \emph {et~al.}(2006)\citenamefont
  {Panaitescu}, \citenamefont {M{\'e}sz{\'a}ros}, \citenamefont {Burrows},
  \citenamefont {Nousek}, \citenamefont {Gehrels}, \citenamefont {O'Brien},\
  and\ \citenamefont {Willingale}}]{Panaitescu&al:MNRAS:2006}%
  \BibitemOpen
  \bibfield  {author} {\bibinfo {author} {\bibfnamefont {A.}~\bibnamefont
  {Panaitescu}}, \bibinfo {author} {\bibfnamefont {P.}~\bibnamefont
  {M{\'e}sz{\'a}ros}}, \bibinfo {author} {\bibfnamefont {D.}~\bibnamefont
  {Burrows}}, \bibinfo {author} {\bibfnamefont {J.}~\bibnamefont {Nousek}},
  \bibinfo {author} {\bibfnamefont {N.}~\bibnamefont {Gehrels}}, \bibinfo
  {author} {\bibfnamefont {P.}~\bibnamefont {O'Brien}},\ and\ \bibinfo
  {author} {\bibfnamefont {R.}~\bibnamefont {Willingale}},\ }\bibfield  {title}
  {\enquote {\bibinfo {title} {Evidence for chromatic {X}-ray light-curve
  breaks in {{\emph{Swift}}} gamma-ray burst afterglows and their theoretical
  implications},}\ }\href {\doibase 10.1111/j.1365-2966.2006.10453.x}
  {\bibfield  {journal} {\bibinfo  {journal} {Mon.\ Not.\ Roy.\ Astron.\ Soc.}\
  }\textbf {\bibinfo {volume} {369}},\ \bibinfo {pages} {2059--2064} (\bibinfo
  {year} {2006})}\BibitemShut {NoStop}%
\bibitem [{\citenamefont {Campana}\ \emph {et~al.}(2007)\citenamefont
  {Campana}, \citenamefont {Guidorzi}, \citenamefont {Tagliaferri},
  \citenamefont {Chincarini}, \citenamefont {Moretti}, \citenamefont
  {Rizzuto},\ and\ \citenamefont {Romano}}]{Campana&al:AA:2007}%
  \BibitemOpen
  \bibfield  {author} {\bibinfo {author} {\bibfnamefont {S.}~\bibnamefont
  {Campana}}, \bibinfo {author} {\bibfnamefont {C.}~\bibnamefont {Guidorzi}},
  \bibinfo {author} {\bibfnamefont {G.}~\bibnamefont {Tagliaferri}}, \bibinfo
  {author} {\bibfnamefont {G.}~\bibnamefont {Chincarini}}, \bibinfo {author}
  {\bibfnamefont {A.}~\bibnamefont {Moretti}}, \bibinfo {author} {\bibfnamefont
  {D.}~\bibnamefont {Rizzuto}},\ and\ \bibinfo {author} {\bibfnamefont
  {P.}~\bibnamefont {Romano}},\ }\bibfield  {title} {\enquote {\bibinfo {title}
  {Are {{\emph{Swift}}} gamma-ray bursts consistent with the {Ghirlanda}
  relation?}}\ }\href {\doibase 10.1051/0004-6361:20066984} {\bibfield
  {journal} {\bibinfo  {journal} {Astron.\ Astrophys.}\ }\textbf {\bibinfo
  {volume} {472}},\ \bibinfo {pages} {395--401} (\bibinfo {year}
  {2007})}\BibitemShut {NoStop}%
\bibitem [{\citenamefont {Liang}\ \emph {et~al.}(2008)\citenamefont {Liang},
  \citenamefont {Xiao}, \citenamefont {Liu},\ and\ \citenamefont
  {Zhang}}]{Liang&al:APJ:2008}%
  \BibitemOpen
  \bibfield  {author} {\bibinfo {author} {\bibfnamefont {Nan}\ \bibnamefont
  {Liang}}, \bibinfo {author} {\bibfnamefont {Wei~Ke}\ \bibnamefont {Xiao}},
  \bibinfo {author} {\bibfnamefont {Yuan}\ \bibnamefont {Liu}},\ and\ \bibinfo
  {author} {\bibfnamefont {Shuang~Nan}\ \bibnamefont {Zhang}},\ }\bibfield
  {title} {\enquote {\bibinfo {title} {A cosmology-independent calibration of
  gamma-ray burst luminosity relations and the {Hubble} diagram},}\ }\href
  {\doibase 10.1086/590903} {\bibfield  {journal} {\bibinfo  {journal}
  {Astrophys.\ J.}\ }\textbf {\bibinfo {volume} {685}},\ \bibinfo {pages}
  {354--360} (\bibinfo {year} {2008})}\BibitemShut {NoStop}%
\bibitem [{\citenamefont {Kann}\ \emph {et~al.}(2010)\citenamefont {Kann},
  \citenamefont {Klose}, \citenamefont {Zhang}, \citenamefont {Malesani},
  \citenamefont {Nakar}, \citenamefont {Pozanenko}, \citenamefont {Wilson},
  \citenamefont {Butler}, \citenamefont {Jakobsson}, \citenamefont {Schulze},
  \citenamefont {Andreev}, \citenamefont {Antonelli}, \citenamefont {Bikmaev},
  \citenamefont {Biryukov}, \citenamefont {B{\"o}ttcher}, \citenamefont
  {Burenin}, \citenamefont {Cer{\'o}n}, \citenamefont {Castro-Tirado},
  \citenamefont {Chincarini}, \citenamefont {Cobb}, \citenamefont {Covino},
  \citenamefont {D'Avanzo}, \citenamefont {D'Elia}, \citenamefont {{Della
  Valle}}, \citenamefont {Postigo}, \citenamefont {Efimov}, \citenamefont
  {Ferrero}, \citenamefont {Fugazza}, \citenamefont {Fynbo}, \citenamefont
  {G{\aa}lfalk}, \citenamefont {Grundahl}, \citenamefont {Gorosabel},
  \citenamefont {Gupta}, \citenamefont {Guziy}, \citenamefont {Hafizov},
  \citenamefont {Hjorth}, \citenamefont {Holhjem}, \citenamefont {Ibrahimov},
  \citenamefont {Im}, \citenamefont {Israel}, \citenamefont {Je{\l}inek},
  \citenamefont {Jensen}, \citenamefont {Karimov}, \citenamefont {Khamitov},
  \citenamefont {Kizilo\u{g}lu}, \citenamefont {Klunko}, \citenamefont
  {Kub{\'a}nek}, \citenamefont {Kutyrev}, \citenamefont {Laursen},
  \citenamefont {Levan}, \citenamefont {Mannucci}, \citenamefont {Martin},
  \citenamefont {Mescheryakov}, \citenamefont {Mirabal}, \citenamefont
  {Norris}, \citenamefont {Ovaldsen}, \citenamefont {Paraficz}, \citenamefont
  {Pavlenko}, \citenamefont {Piranomonte}, \citenamefont {Rossi}, \citenamefont
  {Rumyantsev}, \citenamefont {Salinas}, \citenamefont {Sergeev}, \citenamefont
  {Sharapov}, \citenamefont {Sollerman}, \citenamefont {Stecklum},
  \citenamefont {Stella}, \citenamefont {Tagliaferri}, \citenamefont {Tanvir},
  \citenamefont {Telting}, \citenamefont {V.~Testa}, \citenamefont {Updike},
  \citenamefont {Volnova}, \citenamefont {Watson}, \citenamefont {Wiersema},\
  and\ \citenamefont {Xu}}]{Kann&al:APJ:2010}%
  \BibitemOpen
  \bibfield  {author} {\bibinfo {author} {\bibfnamefont {D.~A.}\ \bibnamefont
  {Kann}}, \bibinfo {author} {\bibfnamefont {S.}~\bibnamefont {Klose}},
  \bibinfo {author} {\bibfnamefont {B.}~\bibnamefont {Zhang}}, \bibinfo
  {author} {\bibfnamefont {D.}~\bibnamefont {Malesani}}, \bibinfo {author}
  {\bibfnamefont {E.}~\bibnamefont {Nakar}}, \bibinfo {author} {\bibfnamefont
  {A.}~\bibnamefont {Pozanenko}}, \bibinfo {author} {\bibfnamefont {A.~C.}\
  \bibnamefont {Wilson}}, \bibinfo {author} {\bibfnamefont {N.~R.}\
  \bibnamefont {Butler}}, \bibinfo {author} {\bibfnamefont {P.}~\bibnamefont
  {Jakobsson}}, \bibinfo {author} {\bibfnamefont {S.}~\bibnamefont {Schulze}},
  \bibinfo {author} {\bibfnamefont {M.}~\bibnamefont {Andreev}}, \bibinfo
  {author} {\bibfnamefont {L.~A.}\ \bibnamefont {Antonelli}}, \bibinfo {author}
  {\bibfnamefont {I.~F.}\ \bibnamefont {Bikmaev}}, \bibinfo {author}
  {\bibfnamefont {V.}~\bibnamefont {Biryukov}}, \bibinfo {author}
  {\bibfnamefont {M.}~\bibnamefont {B{\"o}ttcher}}, \bibinfo {author}
  {\bibfnamefont {R.~A.}\ \bibnamefont {Burenin}}, \bibinfo {author}
  {\bibfnamefont {J.~M.~Castro}\ \bibnamefont {Cer{\'o}n}}, \bibinfo {author}
  {\bibfnamefont {A.~J.}\ \bibnamefont {Castro-Tirado}}, \bibinfo {author}
  {\bibfnamefont {G.}~\bibnamefont {Chincarini}}, \bibinfo {author}
  {\bibfnamefont {B.~E.}\ \bibnamefont {Cobb}}, \bibinfo {author}
  {\bibfnamefont {S.}~\bibnamefont {Covino}}, \bibinfo {author} {\bibfnamefont
  {P.}~\bibnamefont {D'Avanzo}}, \bibinfo {author} {\bibfnamefont
  {V.}~\bibnamefont {D'Elia}}, \bibinfo {author} {\bibfnamefont
  {M.}~\bibnamefont {{Della Valle}}}, \bibinfo {author} {\bibfnamefont
  {A.~de~Ugarte}\ \bibnamefont {Postigo}}, \bibinfo {author} {\bibfnamefont
  {Yu}~\bibnamefont {Efimov}}, \bibinfo {author} {\bibfnamefont
  {P.}~\bibnamefont {Ferrero}}, \bibinfo {author} {\bibfnamefont
  {D.}~\bibnamefont {Fugazza}}, \bibinfo {author} {\bibfnamefont {J.~P.~U.}\
  \bibnamefont {Fynbo}}, \bibinfo {author} {\bibfnamefont {M.}~\bibnamefont
  {G{\aa}lfalk}}, \bibinfo {author} {\bibfnamefont {F.}~\bibnamefont
  {Grundahl}}, \bibinfo {author} {\bibfnamefont {J.}~\bibnamefont {Gorosabel}},
  \bibinfo {author} {\bibfnamefont {S.}~\bibnamefont {Gupta}}, \bibinfo
  {author} {\bibfnamefont {S.}~\bibnamefont {Guziy}}, \bibinfo {author}
  {\bibfnamefont {B.}~\bibnamefont {Hafizov}}, \bibinfo {author} {\bibfnamefont
  {J.}~\bibnamefont {Hjorth}}, \bibinfo {author} {\bibfnamefont
  {K.}~\bibnamefont {Holhjem}}, \bibinfo {author} {\bibfnamefont
  {M.}~\bibnamefont {Ibrahimov}}, \bibinfo {author} {\bibfnamefont
  {M.}~\bibnamefont {Im}}, \bibinfo {author} {\bibfnamefont {G.~L.}\
  \bibnamefont {Israel}}, \bibinfo {author} {\bibfnamefont {M.}~\bibnamefont
  {Je{\l}inek}}, \bibinfo {author} {\bibfnamefont {B.~L.}\ \bibnamefont
  {Jensen}}, \bibinfo {author} {\bibfnamefont {R.}~\bibnamefont {Karimov}},
  \bibinfo {author} {\bibfnamefont {I.~M.}\ \bibnamefont {Khamitov}}, \bibinfo
  {author} {\bibfnamefont {{\"U}.}~\bibnamefont {Kizilo\u{g}lu}}, \bibinfo
  {author} {\bibfnamefont {E.}~\bibnamefont {Klunko}}, \bibinfo {author}
  {\bibfnamefont {P.}~\bibnamefont {Kub{\'a}nek}}, \bibinfo {author}
  {\bibfnamefont {A.~S.}\ \bibnamefont {Kutyrev}}, \bibinfo {author}
  {\bibfnamefont {P.}~\bibnamefont {Laursen}}, \bibinfo {author} {\bibfnamefont
  {A.~J.}\ \bibnamefont {Levan}}, \bibinfo {author} {\bibfnamefont
  {F.}~\bibnamefont {Mannucci}}, \bibinfo {author} {\bibfnamefont {C.~M.}\
  \bibnamefont {Martin}}, \bibinfo {author} {\bibfnamefont {A.}~\bibnamefont
  {Mescheryakov}}, \bibinfo {author} {\bibfnamefont {N.}~\bibnamefont
  {Mirabal}}, \bibinfo {author} {\bibfnamefont {J.~P.}\ \bibnamefont {Norris}},
  \bibinfo {author} {\bibfnamefont {J.-E.}\ \bibnamefont {Ovaldsen}}, \bibinfo
  {author} {\bibfnamefont {D.}~\bibnamefont {Paraficz}}, \bibinfo {author}
  {\bibfnamefont {E.}~\bibnamefont {Pavlenko}}, \bibinfo {author}
  {\bibfnamefont {S.}~\bibnamefont {Piranomonte}}, \bibinfo {author}
  {\bibfnamefont {A.}~\bibnamefont {Rossi}}, \bibinfo {author} {\bibfnamefont
  {V.}~\bibnamefont {Rumyantsev}}, \bibinfo {author} {\bibfnamefont
  {R.}~\bibnamefont {Salinas}}, \bibinfo {author} {\bibfnamefont
  {A.}~\bibnamefont {Sergeev}}, \bibinfo {author} {\bibfnamefont
  {D.}~\bibnamefont {Sharapov}}, \bibinfo {author} {\bibfnamefont
  {J.}~\bibnamefont {Sollerman}}, \bibinfo {author} {\bibfnamefont
  {B.}~\bibnamefont {Stecklum}}, \bibinfo {author} {\bibfnamefont
  {L.}~\bibnamefont {Stella}}, \bibinfo {author} {\bibfnamefont
  {G.}~\bibnamefont {Tagliaferri}}, \bibinfo {author} {\bibfnamefont {N.~R.}\
  \bibnamefont {Tanvir}}, \bibinfo {author} {\bibfnamefont {J.}~\bibnamefont
  {Telting}}, \bibinfo {author} {\bibfnamefont {V.}~\bibnamefont {V.~Testa}},
  \bibinfo {author} {\bibfnamefont {A.~C.}\ \bibnamefont {Updike}}, \bibinfo
  {author} {\bibfnamefont {A.}~\bibnamefont {Volnova}}, \bibinfo {author}
  {\bibfnamefont {D.}~\bibnamefont {Watson}}, \bibinfo {author} {\bibfnamefont
  {K.}~\bibnamefont {Wiersema}},\ and\ \bibinfo {author} {\bibfnamefont
  {D.}~\bibnamefont {Xu}},\ }\bibfield  {title} {\enquote {\bibinfo {title}
  {The afterglows of {{\emph{Swift}}}-era {Gamma-Ray Bursts}. {I}. comparing
  pre-{{\emph{Swift}}} and {{\emph{Swift}}}-era long/soft (type {II}) {GRB}
  optical afterglows},}\ }\href {\doibase 10.1088/0004-637X/720/2/1513}
  {\bibfield  {journal} {\bibinfo  {journal} {Astrophys.\ J.}\ }\textbf
  {\bibinfo {volume} {720}},\ \bibinfo {pages} {1513--1558} (\bibinfo {year}
  {2010})}\BibitemShut {NoStop}%
\bibitem [{\citenamefont {Kann}\ \emph {et~al.}(2011)\citenamefont {Kann},
  \citenamefont {Klose}, \citenamefont {Zhang}, \citenamefont {Covino},
  \citenamefont {Butler}, \citenamefont {Malesani}, \citenamefont {Nakar},
  \citenamefont {Wilson}, \citenamefont {Antonelli}, \citenamefont
  {Chincarini}, \citenamefont {Cobb}, \citenamefont {D'Avanzo}, \citenamefont
  {D'Elia}, \citenamefont {{Della Valle}}, \citenamefont {Ferrero},
  \citenamefont {Fugazza}, \citenamefont {Gorosabel}, \citenamefont {Israel},
  \citenamefont {Mannucci}, \citenamefont {Piranomonte}, \citenamefont
  {Schulze}, \citenamefont {Stella}, \citenamefont {Tagliaferri},\ and\
  \citenamefont {Wiersema}}]{Kann&al:APJ:2011}%
  \BibitemOpen
  \bibfield  {author} {\bibinfo {author} {\bibfnamefont {D.~A.}\ \bibnamefont
  {Kann}}, \bibinfo {author} {\bibfnamefont {S.}~\bibnamefont {Klose}},
  \bibinfo {author} {\bibfnamefont {B.}~\bibnamefont {Zhang}}, \bibinfo
  {author} {\bibfnamefont {S.}~\bibnamefont {Covino}}, \bibinfo {author}
  {\bibfnamefont {N.~R.}\ \bibnamefont {Butler}}, \bibinfo {author}
  {\bibfnamefont {D.}~\bibnamefont {Malesani}}, \bibinfo {author}
  {\bibfnamefont {E.}~\bibnamefont {Nakar}}, \bibinfo {author} {\bibfnamefont
  {A.~C.}\ \bibnamefont {Wilson}}, \bibinfo {author} {\bibfnamefont {L.~A.}\
  \bibnamefont {Antonelli}}, \bibinfo {author} {\bibfnamefont {G.}~\bibnamefont
  {Chincarini}}, \bibinfo {author} {\bibfnamefont {B.~E.}\ \bibnamefont
  {Cobb}}, \bibinfo {author} {\bibfnamefont {P.}~\bibnamefont {D'Avanzo}},
  \bibinfo {author} {\bibfnamefont {V.}~\bibnamefont {D'Elia}}, \bibinfo
  {author} {\bibfnamefont {M.}~\bibnamefont {{Della Valle}}}, \bibinfo {author}
  {\bibfnamefont {P.}~\bibnamefont {Ferrero}}, \bibinfo {author} {\bibfnamefont
  {D.}~\bibnamefont {Fugazza}}, \bibinfo {author} {\bibfnamefont
  {J.}~\bibnamefont {Gorosabel}}, \bibinfo {author} {\bibfnamefont {G.~L.}\
  \bibnamefont {Israel}}, \bibinfo {author} {\bibfnamefont {F.}~\bibnamefont
  {Mannucci}}, \bibinfo {author} {\bibfnamefont {S.}~\bibnamefont
  {Piranomonte}}, \bibinfo {author} {\bibfnamefont {S.}~\bibnamefont
  {Schulze}}, \bibinfo {author} {\bibfnamefont {L.}~\bibnamefont {Stella}},
  \bibinfo {author} {\bibfnamefont {G.}~\bibnamefont {Tagliaferri}},\ and\
  \bibinfo {author} {\bibfnamefont {K.}~\bibnamefont {Wiersema}},\ }\bibfield
  {title} {\enquote {\bibinfo {title} {The afterglows of {{\emph{Swift}}}-era
  {Gamma-Ray Bursts}. {I}. type {I GRB} versus type {II GRB} optical
  afterglows},}\ }\href {\doibase 10.1088/0004-637X/734/2/96} {\bibfield
  {journal} {\bibinfo  {journal} {Astrophys.\ J.}\ }\textbf {\bibinfo {volume}
  {734}},\ \bibinfo {pages} {96} (\bibinfo {year} {2011})}\BibitemShut
  {NoStop}%
\bibitem [{\citenamefont {Amati}\ and\ \citenamefont {{Della
  Valle}}(2013)}]{Amati&DellaValle:IJMPD:2013}%
  \BibitemOpen
  \bibfield  {author} {\bibinfo {author} {\bibfnamefont {Lorenzo}\ \bibnamefont
  {Amati}}\ and\ \bibinfo {author} {\bibfnamefont {Massimo}\ \bibnamefont
  {{Della Valle}}},\ }\bibfield  {title} {\enquote {\bibinfo {title} {Measuring
  cosmological parameters with gamma ray bursts},}\ }\href {\doibase
  10.1142/S0218271813300280} {\bibfield  {journal} {\bibinfo  {journal} {Int.\
  J.~Mod.\ Phys.~D}\ }\textbf {\bibinfo {volume} {22}},\ \bibinfo {pages}
  {1330028} (\bibinfo {year} {2013})}\BibitemShut {NoStop}%
\bibitem [{\citenamefont {{De Pasquale}}\ \emph {et~al.}(2016)\citenamefont
  {{De Pasquale}}, \citenamefont {Page}, \citenamefont {Kann}, \citenamefont
  {Oates}, \citenamefont {Schulze}, \citenamefont {Zhang}, \citenamefont
  {Cano}, \citenamefont {Gendre}, \citenamefont {Malesani}, \citenamefont
  {Rossi}, \citenamefont {Troja}, \citenamefont {Piro}, \citenamefont
  {Bo{\"e}r}, \citenamefont {Stratta},\ and\ \citenamefont
  {Gehrels}}]{DePasquale&al:ARXIV:2016}%
  \BibitemOpen
  \bibfield  {author} {\bibinfo {author} {\bibfnamefont {M.}~\bibnamefont {{De
  Pasquale}}}, \bibinfo {author} {\bibfnamefont {M.~J.}\ \bibnamefont {Page}},
  \bibinfo {author} {\bibfnamefont {D.~A.}\ \bibnamefont {Kann}}, \bibinfo
  {author} {\bibfnamefont {S.~R.}\ \bibnamefont {Oates}}, \bibinfo {author}
  {\bibfnamefont {S.}~\bibnamefont {Schulze}}, \bibinfo {author} {\bibfnamefont
  {B.}~\bibnamefont {Zhang}}, \bibinfo {author} {\bibfnamefont
  {Z.}~\bibnamefont {Cano}}, \bibinfo {author} {\bibfnamefont {B.}~\bibnamefont
  {Gendre}}, \bibinfo {author} {\bibfnamefont {D.}~\bibnamefont {Malesani}},
  \bibinfo {author} {\bibfnamefont {A.}~\bibnamefont {Rossi}}, \bibinfo
  {author} {\bibfnamefont {E.}~\bibnamefont {Troja}}, \bibinfo {author}
  {\bibfnamefont {L.}~\bibnamefont {Piro}}, \bibinfo {author} {\bibfnamefont
  {M.}~\bibnamefont {Bo{\"e}r}}, \bibinfo {author} {\bibfnamefont
  {G.}~\bibnamefont {Stratta}},\ and\ \bibinfo {author} {\bibfnamefont
  {N.}~\bibnamefont {Gehrels}},\ }\href {http://arxiv.org/abs/1602.04158}
  {\enquote {\bibinfo {title} {The 80 {Ms} follow-up of the {X}-ray afterglow
  of {GRB 130427A} challenges the standard forward shock model},}\ } (\bibinfo
  {year} {2016}),\ \Eprint {http://arxiv.org/abs/1602.04158}
  {arXiv.org:1602.04158 [astro-ph.HE]} \BibitemShut {NoStop}%
\bibitem [{\citenamefont {Sari}\ \emph {et~al.}(1999)\citenamefont {Sari},
  \citenamefont {Piran},\ and\ \citenamefont {Halpern}}]{Sari&al:APJL:1999}%
  \BibitemOpen
  \bibfield  {author} {\bibinfo {author} {\bibfnamefont {{Re'em}}\ \bibnamefont
  {Sari}}, \bibinfo {author} {\bibfnamefont {Tsvi}\ \bibnamefont {Piran}}, \
  and\ \bibinfo {author} {\bibfnamefont {J.~P.}\ \bibnamefont {Halpern}},\
  }\bibfield  {title} {\enquote {\bibinfo {title} {Jets in {Gamma-Ray
  Bursts}},}\ }\href {\doibase 10.1086/312109} {\bibfield  {journal} {\bibinfo
  {journal} {Astrophys.\ J.~Lett.}\ }\textbf {\bibinfo {volume} {519}},\
  \bibinfo {pages} {L17--L20} (\bibinfo {year} {1999})}\BibitemShut {NoStop}%
\bibitem [{\citenamefont {Frail}\ \emph {et~al.}(2001)\citenamefont {Frail},
  \citenamefont {Kulkarni}, \citenamefont {Sari}, \citenamefont {Djorgovski},
  \citenamefont {Bloom}, \citenamefont {Galama}, \citenamefont {Reichart},
  \citenamefont {Berger}, \citenamefont {Harrison}, \citenamefont {Price},
  \citenamefont {Yost}, \citenamefont {Diercks}, \citenamefont {Goodrich},\
  and\ \citenamefont {Chaffee}}]{Frail&al:APJL:2001}%
  \BibitemOpen
  \bibfield  {author} {\bibinfo {author} {\bibfnamefont {D.~A.}\ \bibnamefont
  {Frail}}, \bibinfo {author} {\bibfnamefont {S.~R.}\ \bibnamefont {Kulkarni}},
  \bibinfo {author} {\bibfnamefont {R.}~\bibnamefont {Sari}}, \bibinfo {author}
  {\bibfnamefont {S.~G.}\ \bibnamefont {Djorgovski}}, \bibinfo {author}
  {\bibfnamefont {J.~S.}\ \bibnamefont {Bloom}}, \bibinfo {author}
  {\bibfnamefont {T.~J.}\ \bibnamefont {Galama}}, \bibinfo {author}
  {\bibfnamefont {D.~E.}\ \bibnamefont {Reichart}}, \bibinfo {author}
  {\bibfnamefont {E.}~\bibnamefont {Berger}}, \bibinfo {author} {\bibfnamefont
  {F.~A.}\ \bibnamefont {Harrison}}, \bibinfo {author} {\bibfnamefont {P.~A.}\
  \bibnamefont {Price}}, \bibinfo {author} {\bibfnamefont {S.~A.}\ \bibnamefont
  {Yost}}, \bibinfo {author} {\bibfnamefont {A.}~\bibnamefont {Diercks}},
  \bibinfo {author} {\bibfnamefont {R.~W.}\ \bibnamefont {Goodrich}},\ and\
  \bibinfo {author} {\bibfnamefont {F.}~\bibnamefont {Chaffee}},\ }\bibfield
  {title} {\enquote {\bibinfo {title} {Beaming in gamma-ray bursts: Evidence
  for a standard energy reservoir},}\ }\href {\doibase 10.1086/338119}
  {\bibfield  {journal} {\bibinfo  {journal} {Astrophys.\ J.~Lett.}\ }\textbf
  {\bibinfo {volume} {562}},\ \bibinfo {pages} {L55--L58} (\bibinfo {year}
  {2001})}\BibitemShut {NoStop}%
\bibitem [{\citenamefont {Ghirlanda}\ \emph
  {et~al.}(2007{\natexlab{a}})\citenamefont {Ghirlanda}, \citenamefont
  {Bosnjak}, \citenamefont {Ghisellini}, \citenamefont {Tavecchio},\ and\
  \citenamefont {Firmani}}]{Ghirlanda&al:MNRAS:2007}%
  \BibitemOpen
  \bibfield  {author} {\bibinfo {author} {\bibfnamefont {G.}~\bibnamefont
  {Ghirlanda}}, \bibinfo {author} {\bibfnamefont {Z.}~\bibnamefont {Bosnjak}},
  \bibinfo {author} {\bibfnamefont {G.}~\bibnamefont {Ghisellini}}, \bibinfo
  {author} {\bibfnamefont {F.}~\bibnamefont {Tavecchio}},\ and\ \bibinfo
  {author} {\bibfnamefont {C.}~\bibnamefont {Firmani}},\ }\bibfield  {title}
  {\enquote {\bibinfo {title} {Blackbody components in gamma-ray bursts
  spectra?}}\ }\href {\doibase 10.1111/j.1365-2966.2007.11890.x} {\bibfield
  {journal} {\bibinfo  {journal} {Mon.\ Not.\ Roy.\ Astron.\ Soc.}\ }\textbf
  {\bibinfo {volume} {379}},\ \bibinfo {pages} {73--85} (\bibinfo {year}
  {2007}{\natexlab{a}})}\BibitemShut {NoStop}%
\bibitem [{\citenamefont {Ghirlanda}\ \emph
  {et~al.}(2007{\natexlab{b}})\citenamefont {Ghirlanda}, \citenamefont {Nava},
  \citenamefont {Ghisellini},\ and\ \citenamefont
  {Firmani}}]{Ghirlanda&al:AA:2007}%
  \BibitemOpen
  \bibfield  {author} {\bibinfo {author} {\bibfnamefont {G.}~\bibnamefont
  {Ghirlanda}}, \bibinfo {author} {\bibfnamefont {L.}~\bibnamefont {Nava}},
  \bibinfo {author} {\bibfnamefont {G.}~\bibnamefont {Ghisellini}},\ and\
  \bibinfo {author} {\bibfnamefont {C.}~\bibnamefont {Firmani}},\ }\bibfield
  {title} {\enquote {\bibinfo {title} {Confirming the {$\gamma$}-ray burst
  spectral-energy correlations in the era of multiple time breaks},}\ }\href
  {\doibase 10.1051/0004-6361:20077119} {\bibfield  {journal} {\bibinfo
  {journal} {Astron.\ Astrophys.}\ }\textbf {\bibinfo {volume} {466}},\
  \bibinfo {pages} {127--136} (\bibinfo {year}
  {2007}{\natexlab{b}})}\BibitemShut {NoStop}%
\bibitem [{\citenamefont {{Della Valle}}(2011)}]{DellaValle:IJMPD:2011}%
  \BibitemOpen
  \bibfield  {author} {\bibinfo {author} {\bibfnamefont {Massimo}\ \bibnamefont
  {{Della Valle}}},\ }\bibfield  {title} {\enquote {\bibinfo {title}
  {Supernovae and gamma-ray bursts: a decade of observations},}\ }\href
  {\doibase 10.1142/S0218271811019827} {\bibfield  {journal} {\bibinfo
  {journal} {Int.\ J.~Mod.\ Phys.~D}\ }\textbf {\bibinfo {volume} {20}},\
  \bibinfo {pages} {1745--1754} (\bibinfo {year} {2011})}\BibitemShut {NoStop}%
\bibitem [{\citenamefont {Guetta}\ and\ \citenamefont {{Della
  Valle}}(2007)}]{Guetta&DellaValle:APJL:2007}%
  \BibitemOpen
  \bibfield  {author} {\bibinfo {author} {\bibfnamefont {Dafne}\ \bibnamefont
  {Guetta}}\ and\ \bibinfo {author} {\bibfnamefont {Massimo}\ \bibnamefont
  {{Della Valle}}},\ }\bibfield  {title} {\enquote {\bibinfo {title} {On the
  rates of gamma-ray bursts and type {I}b/c supernovae},}\ }\href {\doibase
  10.1086/511417} {\bibfield  {journal} {\bibinfo  {journal} {Astrophys.\
  J.~Lett.}\ }\textbf {\bibinfo {volume} {657}},\ \bibinfo {pages} {L73--L76}
  (\bibinfo {year} {2007})}\BibitemShut {NoStop}%
\bibitem [{\citenamefont {Chandra}\ and\ \citenamefont
  {Frail}(2012)}]{Chandra&Frail:APJ:2012}%
  \BibitemOpen
  \bibfield  {author} {\bibinfo {author} {\bibfnamefont {Poonam}\ \bibnamefont
  {Chandra}}\ and\ \bibinfo {author} {\bibfnamefont {Dale~A.}\ \bibnamefont
  {Frail}},\ }\bibfield  {title} {\enquote {\bibinfo {title} {A radio-selected
  sample of gamma-ray burst afterglows},}\ }\href {\doibase
  10.1088/0004-637X/746/2/156} {\bibfield  {journal} {\bibinfo  {journal}
  {Astrophys.\ J.}\ }\textbf {\bibinfo {volume} {746}},\ \bibinfo {pages} {156}
  (\bibinfo {year} {2012})}\BibitemShut {NoStop}%
\bibitem [{\citenamefont {Guetta}\ and\ \citenamefont
  {Piran}(2007)}]{Guetta&Piran:JCAP:2007}%
  \BibitemOpen
  \bibfield  {author} {\bibinfo {author} {\bibfnamefont {Dafne}\ \bibnamefont
  {Guetta}}\ and\ \bibinfo {author} {\bibfnamefont {Tsvi}\ \bibnamefont
  {Piran}},\ }\bibfield  {title} {\enquote {\bibinfo {title} {Do long duration
  gamma ray bursts follow star formation?}}\ }\href {\doibase
  10.1088/1475-7516/2007/07/003} {\bibfield  {journal} {\bibinfo  {journal}
  {J.~Cosm.\ Astropart.\ Phys.}\ }\textbf {\bibinfo {volume} {2007}},\ \bibinfo
  {pages} {003} (\bibinfo {year} {2007})}\BibitemShut {NoStop}%
\bibitem [{\citenamefont {Woosley}(1993)}]{Woosley:APJ:1993}%
  \BibitemOpen
  \bibfield  {author} {\bibinfo {author} {\bibfnamefont {S.~E.}\ \bibnamefont
  {Woosley}},\ }\bibfield  {title} {\enquote {\bibinfo {title} {Gamma-ray
  bursts from stellar mass accretion disks around black holes},}\ }\href@noop
  {} {\bibfield  {journal} {\bibinfo  {journal} {Astrophys.\ J.}\ }\textbf
  {\bibinfo {volume} {405}},\ \bibinfo {pages} {273--277} (\bibinfo {year}
  {1993})}\BibitemShut {NoStop}%
\bibitem [{\citenamefont {Woosley}\ and\ \citenamefont
  {Bloom}(2006)}]{Woosley&Bloom:ARAA:2006}%
  \BibitemOpen
  \bibfield  {author} {\bibinfo {author} {\bibfnamefont {S.~E.}\ \bibnamefont
  {Woosley}}\ and\ \bibinfo {author} {\bibfnamefont {J.~S.}\ \bibnamefont
  {Bloom}},\ }\bibfield  {title} {\enquote {\bibinfo {title} {The
  supernova--gamma-ray burst connection},}\ }\href {\doibase
  10.1146/annurev.astro.43.072103.150558} {\bibfield  {journal} {\bibinfo
  {journal} {Ann.\ Rev.\ Astron.\ Astrophys.}\ }\textbf {\bibinfo {volume}
  {44}},\ \bibinfo {pages} {507--556} (\bibinfo {year} {2006})}\BibitemShut
  {NoStop}%
\bibitem [{\citenamefont {Lyutikov}\ and\ \citenamefont
  {Usov}(2000)}]{Lyutikov&Usov:APJL:2000}%
  \BibitemOpen
  \bibfield  {author} {\bibinfo {author} {\bibfnamefont {Maxim}\ \bibnamefont
  {Lyutikov}}\ and\ \bibinfo {author} {\bibfnamefont {Vladimir~V.}\
  \bibnamefont {Usov}},\ }\bibfield  {title} {\enquote {\bibinfo {title}
  {Precursors of gamma-ray bursts: A clue to the burster's nature},}\ }\href
  {\doibase 10.1086/317278} {\bibfield  {journal} {\bibinfo  {journal}
  {Astrophys.\ J.~Lett.}\ }\textbf {\bibinfo {volume} {543}} (\bibinfo {year}
  {2000}),\ 10.1086/317278}\BibitemShut {NoStop}%
\bibitem [{\citenamefont {Zhang}\ \emph {et~al.}(2004)\citenamefont {Zhang},
  \citenamefont {Woosley},\ and\ \citenamefont {Heger}}]{Zhang&al:APJ:2004}%
  \BibitemOpen
  \bibfield  {author} {\bibinfo {author} {\bibfnamefont {Weiqun}\ \bibnamefont
  {Zhang}}, \bibinfo {author} {\bibfnamefont {S.~E.}\ \bibnamefont {Woosley}},
  \ and\ \bibinfo {author} {\bibfnamefont {A.}~\bibnamefont {Heger}},\
  }\bibfield  {title} {\enquote {\bibinfo {title} {The propagation and eruption
  of relativistic jets from the stellar progenitors of gamma-ray bursts},}\
  }\href {\doibase 10.1086/386300} {\bibfield  {journal} {\bibinfo  {journal}
  {Astrophys.\ J.}\ }\textbf {\bibinfo {volume} {608}},\ \bibinfo {pages}
  {365--377} (\bibinfo {year} {2004})}\BibitemShut {NoStop}%
\bibitem [{\citenamefont {Metzger}\ \emph {et~al.}(2010)\citenamefont
  {Metzger}, \citenamefont {Mart{\'\i}nez-Pinedo}, \citenamefont {Darbha},
  \citenamefont {Quataert}, \citenamefont {Arcones}, \citenamefont {Kasen},
  \citenamefont {Thomas}, \citenamefont {Nugent}, \citenamefont {Panov},\ and\
  \citenamefont {Zinner}}]{Metzger&al:MNRAS:2010}%
  \BibitemOpen
  \bibfield  {author} {\bibinfo {author} {\bibfnamefont {B.~D.}\ \bibnamefont
  {Metzger}}, \bibinfo {author} {\bibfnamefont {G.}~\bibnamefont
  {Mart{\'\i}nez-Pinedo}}, \bibinfo {author} {\bibfnamefont {S.}~\bibnamefont
  {Darbha}}, \bibinfo {author} {\bibfnamefont {E.}~\bibnamefont {Quataert}},
  \bibinfo {author} {\bibfnamefont {A.}~\bibnamefont {Arcones}}, \bibinfo
  {author} {\bibfnamefont {D.}~\bibnamefont {Kasen}}, \bibinfo {author}
  {\bibfnamefont {R.}~\bibnamefont {Thomas}}, \bibinfo {author} {\bibfnamefont
  {P.}~\bibnamefont {Nugent}}, \bibinfo {author} {\bibfnamefont {I.~V.}\
  \bibnamefont {Panov}},\ and\ \bibinfo {author} {\bibfnamefont {N.~T.}\
  \bibnamefont {Zinner}},\ }\bibfield  {title} {\enquote {\bibinfo {title}
  {Electromagnetic counterparts of compact object mergers powered by the
  radioactive decay of {$r$}-process nuclei},}\ }\href {\doibase
  10.1111/j.1365-2966.2010.16864.x} {\bibfield  {journal} {\bibinfo  {journal}
  {Mon.\ Not.\ Roy.\ Astron.\ Soc.}\ }\textbf {\bibinfo {volume} {406}},\
  \bibinfo {pages} {2650--2662} (\bibinfo {year} {2010})}\BibitemShut {NoStop}%
\bibitem [{\citenamefont {Amati}\ \emph {et~al.}(2007)\citenamefont {Amati},
  \citenamefont {{Della Valle}}, \citenamefont {Frontera}, \citenamefont
  {Malesani}, \citenamefont {Guidorzi}, \citenamefont {Montanari},\ and\
  \citenamefont {Pian}}]{Amati&al:AA:2007}%
  \BibitemOpen
  \bibfield  {author} {\bibinfo {author} {\bibfnamefont {L.}~\bibnamefont
  {Amati}}, \bibinfo {author} {\bibfnamefont {M.}~\bibnamefont {{Della
  Valle}}}, \bibinfo {author} {\bibfnamefont {F.}~\bibnamefont {Frontera}},
  \bibinfo {author} {\bibfnamefont {D.}~\bibnamefont {Malesani}}, \bibinfo
  {author} {\bibfnamefont {C.}~\bibnamefont {Guidorzi}}, \bibinfo {author}
  {\bibfnamefont {E.}~\bibnamefont {Montanari}},\ and\ \bibinfo {author}
  {\bibfnamefont {E.}~\bibnamefont {Pian}},\ }\bibfield  {title} {\enquote
  {\bibinfo {title} {On the consistency of peculiar {GRBs}~060218 and 060614
  with the {$E_\text{p,i}$} -- {$E_\text{iso}$} correlation},}\ }\href
  {\doibase 10.1051/0004-6361:20065994} {\bibfield  {journal} {\bibinfo
  {journal} {Astron.\ Astrophys.}\ }\textbf {\bibinfo {volume} {463}},\
  \bibinfo {pages} {913--919} (\bibinfo {year} {2007})}\BibitemShut {NoStop}%
\bibitem [{\citenamefont {Capozziello}\ and\ \citenamefont
  {Lambiase}(2015)}]{Capozziello&Lambiase:PLB:2015}%
  \BibitemOpen
  \bibfield  {author} {\bibinfo {author} {\bibfnamefont {S.}~\bibnamefont
  {Capozziello}}\ and\ \bibinfo {author} {\bibfnamefont {G.}~\bibnamefont
  {Lambiase}},\ }\bibfield  {title} {\enquote {\bibinfo {title} {The emission
  of gamma ray bursts as a test-bed for modified gravity},}\ }\href {\doibase
  10.1016/j.physletb.2015.09.048} {\bibfield  {journal} {\bibinfo  {journal}
  {Phys.\ Lett.\ B}\ }\textbf {\bibinfo {volume} {750}},\ \bibinfo {pages}
  {344--347} (\bibinfo {year} {2015})}\BibitemShut {NoStop}%
\bibitem [{\citenamefont {Bhat}\ \emph {et~al.}(1985)\citenamefont {Bhat},
  \citenamefont {Dhurandhar},\ and\ \citenamefont
  {Dadhich}}]{Bhat&al:JAA:1985}%
  \BibitemOpen
  \bibfield  {author} {\bibinfo {author} {\bibfnamefont {Manjiri}\ \bibnamefont
  {Bhat}}, \bibinfo {author} {\bibfnamefont {Sanjeev}\ \bibnamefont
  {Dhurandhar}},\ and\ \bibinfo {author} {\bibfnamefont {Naresh}\ \bibnamefont
  {Dadhich}},\ }\bibfield  {title} {\enquote {\bibinfo {title} {Energetics of
  the {Kerr-Newman} black hole by the {Penrose} process},}\ }\href {\doibase
  10.1007/BF02715080} {\bibfield  {journal} {\bibinfo  {journal}
  {J.~Astrophys.\ Astron.}\ }\textbf {\bibinfo {volume} {6}},\ \bibinfo {pages}
  {85--100} (\bibinfo {year} {1985})}\BibitemShut {NoStop}%
\bibitem [{\citenamefont {B{\"o}hmer}\ and\ \citenamefont
  {Harko}(2007)}]{Bohmer&Harko:EPJC:2007}%
  \BibitemOpen
  \bibfield  {author} {\bibinfo {author} {\bibfnamefont {C.~G.}\ \bibnamefont
  {B{\"o}hmer}}\ and\ \bibinfo {author} {\bibfnamefont {T.}~\bibnamefont
  {Harko}},\ }\bibfield  {title} {\enquote {\bibinfo {title} {Dark energy as a
  massive vector field},}\ }\href {\doibase 10.1140/epjc/s10052-007-0210-1}
  {\bibfield  {journal} {\bibinfo  {journal} {Eur.\ Phys.~J.~C}\ }\textbf
  {\bibinfo {volume} {50}},\ \bibinfo {pages} {423--429} (\bibinfo {year}
  {2007})}\BibitemShut {NoStop}%
\bibitem [{\citenamefont {Schwinger}(1962)}]{Schwinger:PR:1962}%
  \BibitemOpen
  \bibfield  {author} {\bibinfo {author} {\bibfnamefont {Julian}\ \bibnamefont
  {Schwinger}},\ }\bibfield  {title} {\enquote {\bibinfo {title} {Gauge
  invariance and mass},}\ }\href {\doibase 10.1103/PhysRev.125.397} {\bibfield
  {journal} {\bibinfo  {journal} {Phys.\ Rev.}\ }\textbf {\bibinfo {volume}
  {125}},\ \bibinfo {pages} {397--398} (\bibinfo {year} {1962})}\BibitemShut
  {NoStop}%
\bibitem [{\citenamefont {Anderson}(1963)}]{Anderson:PR:1963}%
  \BibitemOpen
  \bibfield  {author} {\bibinfo {author} {\bibfnamefont {P.~W.}\ \bibnamefont
  {Anderson}},\ }\bibfield  {title} {\enquote {\bibinfo {title} {Plasmons,
  gauge invariance, and mass},}\ }\href {\doibase 10.1103/PhysRev.130.439}
  {\bibfield  {journal} {\bibinfo  {journal} {Phys.\ Rev.}\ }\textbf {\bibinfo
  {volume} {130}},\ \bibinfo {pages} {439--442} (\bibinfo {year}
  {1963})}\BibitemShut {NoStop}%
\bibitem [{\citenamefont {Mendon\c{c}a}(2001)}]{Mendonca:Book:2001}%
  \BibitemOpen
  \bibfield  {author} {\bibinfo {author} {\bibfnamefont {Jose\'{e}~Tito}\
  \bibnamefont {Mendon\c{c}a}},\ }\href@noop {} {\emph {\bibinfo {title}
  {Theory of Photon Acceleration}}}\ (\bibinfo  {publisher} {IOP Publishing},\
  \bibinfo {address} {Bristol, UK},\ \bibinfo {year} {2001})\BibitemShut
  {NoStop}%
\bibitem [{\citenamefont {Tamburini}\ \emph {et~al.}(2010)\citenamefont
  {Tamburini}, \citenamefont {Sponselli}, \citenamefont {Thid\'{e}},\ and\
  \citenamefont {Mendon\c{c}a}}]{Tamburini&al:EPL:2010}%
  \BibitemOpen
  \bibfield  {author} {\bibinfo {author} {\bibfnamefont {F.}~\bibnamefont
  {Tamburini}}, \bibinfo {author} {\bibfnamefont {A.}~\bibnamefont
  {Sponselli}}, \bibinfo {author} {\bibfnamefont {B.}~\bibnamefont
  {Thid\'{e}}},\ and\ \bibinfo {author} {\bibfnamefont {J.~T.}\ \bibnamefont
  {Mendon\c{c}a}},\ }\bibfield  {title} {\enquote {\bibinfo {title} {Photon
  orbital angular momentum and mass in a plasma vortex},}\ }\href {\doibase
  10.1209/0295-5075/90/45001} {\bibfield  {journal} {\bibinfo  {journal}
  {Europhys.\ Lett.}\ }\textbf {\bibinfo {volume} {90}},\ \bibinfo {pages}
  {45001} (\bibinfo {year} {2010})}\BibitemShut {NoStop}%
\bibitem [{\citenamefont {Bei}\ \emph {et~al.}(2004)\citenamefont {Bei},
  \citenamefont {Shi},\ and\ \citenamefont {Liu}}]{Bei&al:IJTP:2004}%
  \BibitemOpen
  \bibfield  {author} {\bibinfo {author} {\bibfnamefont {Xiaomin}\ \bibnamefont
  {Bei}}, \bibinfo {author} {\bibfnamefont {Changsheng}\ \bibnamefont {Shi}}, \
  and\ \bibinfo {author} {\bibfnamefont {Zhongzhu}\ \bibnamefont {Liu}},\
  }\bibfield  {title} {\enquote {\bibinfo {title} {Proca effect in
  {Kerr-Newman} metric},}\ }\href {\doibase 10.1023/B:IJTP.0000048638.90043.97}
  {\bibfield  {journal} {\bibinfo  {journal} {Int.\ J.~Theor.\ Phys.}\ }\textbf
  {\bibinfo {volume} {43}},\ \bibinfo {pages} {1555--1560} (\bibinfo {year}
  {2004})}\BibitemShut {NoStop}%
\bibitem [{\citenamefont {Shi}\ and\ \citenamefont
  {Liu}(2005)}]{Shi&Liu:IJTP:2005}%
  \BibitemOpen
  \bibfield  {author} {\bibinfo {author} {\bibfnamefont {Changsheng}\
  \bibnamefont {Shi}}\ and\ \bibinfo {author} {\bibfnamefont {Zhongzhu}\
  \bibnamefont {Liu}},\ }\bibfield  {title} {\enquote {\bibinfo {title} {Proca
  effect in {Reissner-Nordstrom} de {Sitter} metric},}\ }\href {\doibase
  10.1007/s10773-005-2992-y} {\bibfield  {journal} {\bibinfo  {journal} {Int.\
  J.~Theor.\ Phys.}\ }\textbf {\bibinfo {volume} {44}},\ \bibinfo {pages}
  {303--308} (\bibinfo {year} {2005})}\BibitemShut {NoStop}%
\bibitem [{\citenamefont {Pani}\ \emph
  {et~al.}(2012{\natexlab{a}})\citenamefont {Pani}, \citenamefont {Cardoso},
  \citenamefont {Gualtieri}, \citenamefont {Berti},\ and\ \citenamefont
  {Ishibashi}}]{Pani&al:PRL:2012}%
  \BibitemOpen
  \bibfield  {author} {\bibinfo {author} {\bibfnamefont {Paolo}\ \bibnamefont
  {Pani}}, \bibinfo {author} {\bibfnamefont {Vitor}\ \bibnamefont {Cardoso}},
  \bibinfo {author} {\bibfnamefont {Leonardo}\ \bibnamefont {Gualtieri}},
  \bibinfo {author} {\bibfnamefont {Emanuele}\ \bibnamefont {Berti}},\ and\
  \bibinfo {author} {\bibfnamefont {Akihiro}\ \bibnamefont {Ishibashi}},\
  }\bibfield  {title} {\enquote {\bibinfo {title} {Black-hole bombs and
  photon-mass bounds},}\ }\href {\doibase 10.1103/PhysRevLett.109.131102}
  {\bibfield  {journal} {\bibinfo  {journal} {Phys.\ Rev.\ Lett.}\ }\textbf
  {\bibinfo {volume} {109}},\ \bibinfo {pages} {131102} (\bibinfo {year}
  {2012}{\natexlab{a}})}\BibitemShut {NoStop}%
\bibitem [{\citenamefont {Pani}\ \emph
  {et~al.}(2012{\natexlab{b}})\citenamefont {Pani}, \citenamefont {Cardoso},
  \citenamefont {Gualtieri}, \citenamefont {Berti},\ and\ \citenamefont
  {Ishibashi}}]{Pani&al:PRD:2012}%
  \BibitemOpen
  \bibfield  {author} {\bibinfo {author} {\bibfnamefont {Paolo}\ \bibnamefont
  {Pani}}, \bibinfo {author} {\bibfnamefont {Vitor}\ \bibnamefont {Cardoso}},
  \bibinfo {author} {\bibfnamefont {Leonardo}\ \bibnamefont {Gualtieri}},
  \bibinfo {author} {\bibfnamefont {Emanuele}\ \bibnamefont {Berti}},\ and\
  \bibinfo {author} {\bibfnamefont {Akihiro}\ \bibnamefont {Ishibashi}},\
  }\bibfield  {title} {\enquote {\bibinfo {title} {Perturbations of slowly
  rotating black holes: Massive vector fields in the {Kerr} metric},}\ }\href
  {\doibase 10.1103/PhysRevD.86.104017} {\bibfield  {journal} {\bibinfo
  {journal} {Phys.\ Rev.\ D}\ }\textbf {\bibinfo {volume} {86}},\ \bibinfo
  {pages} {104017} (\bibinfo {year} {2012}{\natexlab{b}})}\BibitemShut
  {NoStop}%
\bibitem [{\citenamefont {Tamburini}\ and\ \citenamefont
  {Thid\'{e}}(2011)}]{Tamburini&Thide:EPL:2011}%
  \BibitemOpen
  \bibfield  {author} {\bibinfo {author} {\bibfnamefont {F.}~\bibnamefont
  {Tamburini}}\ and\ \bibinfo {author} {\bibfnamefont {B.}~\bibnamefont
  {Thid\'{e}}},\ }\bibfield  {title} {\enquote {\bibinfo {title} {Storming
  {Majorana's Tower} with {OAM} states of light in a plasma},}\ }\href
  {\doibase 10.1209/0295-5075/96/64005} {\bibfield  {journal} {\bibinfo
  {journal} {Europhys.\ Lett.}\ }\textbf {\bibinfo {volume} {96}},\ \bibinfo
  {pages} {64005} (\bibinfo {year} {2011})}\BibitemShut {NoStop}%
\bibitem [{\citenamefont {Capozziello}\ and\ \citenamefont {{De
  Laurentis}}(2012)}]{Capozziello&DeLaurentis:APB:2012}%
  \BibitemOpen
  \bibfield  {author} {\bibinfo {author} {\bibfnamefont {S.}~\bibnamefont
  {Capozziello}}\ and\ \bibinfo {author} {\bibfnamefont {M.}~\bibnamefont {{De
  Laurentis}}},\ }\bibfield  {title} {\enquote {\bibinfo {title} {The dark
  matter problem from {$f(R)$} gravity viewpoint},}\ }\href {\doibase
  10.1002/andp.201200109} {\bibfield  {journal} {\bibinfo  {journal} {Ann.\
  Phys. (Berlin)}\ }\textbf {\bibinfo {volume} {524}},\ \bibinfo {pages}
  {545--578} (\bibinfo {year} {2012})}\BibitemShut {NoStop}%
\bibitem [{\citenamefont {Capozziello}\ and\ \citenamefont {{De
  Laurentis}}(2011)}]{Capozziello&DeLaurentis:PHR:2011}%
  \BibitemOpen
  \bibfield  {author} {\bibinfo {author} {\bibfnamefont {Salvatore}\
  \bibnamefont {Capozziello}}\ and\ \bibinfo {author} {\bibfnamefont
  {Mariafelicia}\ \bibnamefont {{De Laurentis}}},\ }\bibfield  {title}
  {\enquote {\bibinfo {title} {Extended theories of gravity},}\ }\href
  {\doibase 10.1016/j.physrep.2011.09.003} {\bibfield  {journal} {\bibinfo
  {journal} {Phys.\ Rep.}\ }\textbf {\bibinfo {volume} {509}},\ \bibinfo
  {pages} {167--321} (\bibinfo {year} {2011})}\BibitemShut {NoStop}%
\bibitem [{\citenamefont {Shapiro}\ \emph {et~al.}(2005)\citenamefont
  {Shapiro}, \citenamefont {Stanev},\ and\ \citenamefont
  {Wefel}}]{Shapiro&al:NATO:2005}%
  \BibitemOpen
  \bibinfo {editor} {\bibfnamefont {Maurice~M.}\ \bibnamefont {Shapiro}},
  \bibinfo {editor} {\bibfnamefont {Todor}\ \bibnamefont {Stanev}},\ and\
  \bibinfo {editor} {\bibfnamefont {John~P.}\ \bibnamefont {Wefel}},\ eds.,\
  \href@noop {} {\emph {\bibinfo {title} {Neutrinos and Explosive Events in the
  Universe}}},\ Proceedings of the NATO Advanced Study Institute\ (\bibinfo
  {publisher} {Springer Netherlands},\ \bibinfo {year} {2005})\ pp.\ \bibinfo
  {pages} {1--424}\BibitemShut {NoStop}%
\bibitem [{\citenamefont {Abbott}\ \emph
  {et~al.}(2016{\natexlab{a}})\citenamefont {Abbott} \emph
  {et~al.}}]{Abbott&al_LIGO:ARXIV:2016}%
  \BibitemOpen
  \bibfield  {author} {\bibinfo {author} {\bibfnamefont {B.~P.}\ \bibnamefont
  {Abbott}} \emph {et~al.} (\bibinfo {collaboration} {LIGO Scientific
  Collaboration and Virgo Collaboration}),\ }\href
  {http://arxiv.org/abs/1602.03841} {\enquote {\bibinfo {title} {Tests of
  general relativity with {GW150914}},}\ } (\bibinfo {year}
  {2016}{\natexlab{a}}),\ \Eprint {http://arxiv.org/abs/1602.03841}
  {arXiv.org:1602.03841 [gr-qc]} \BibitemShut {NoStop}%
\bibitem [{\citenamefont {Abbott}\ \emph
  {et~al.}(2016{\natexlab{b}})\citenamefont {Abbott} \emph
  {et~al.}}]{Abbott&al_LIGO:PRL:2016}%
  \BibitemOpen
  \bibfield  {author} {\bibinfo {author} {\bibfnamefont {B.~P.}\ \bibnamefont
  {Abbott}} \emph {et~al.} (\bibinfo {collaboration} {LIGO Scientific
  Collaboration and Virgo Collaboration}),\ }\bibfield  {title} {\enquote
  {\bibinfo {title} {Observation of gravitational waves from a binary black
  hole merger},}\ }\href {\doibase 10.1103/PhysRevLett.116.061102} {\bibfield
  {journal} {\bibinfo  {journal} {Phys.\ Rev.\ Lett.}\ }\textbf {\bibinfo
  {volume} {116}},\ \bibinfo {pages} {061102} (\bibinfo {year}
  {2016}{\natexlab{b}})}\BibitemShut {NoStop}%
\bibitem [{\citenamefont {Loeb}(2016)}]{Loeb:AJP:2016}%
  \BibitemOpen
  \bibfield  {author} {\bibinfo {author} {\bibfnamefont {Abraham}\ \bibnamefont
  {Loeb}},\ }\bibfield  {title} {\enquote {\bibinfo {title} {Electromagnetic
  counterparts to black hole mergers detected by {LIGO}},}\ }\href {\doibase
  10.3847/2041-8205/819/2/L21} {\bibfield  {journal} {\bibinfo  {journal}
  {Astrophys.\ J.}\ }\textbf {\bibinfo {volume} {819}},\ \bibinfo {pages} {L21}
  (\bibinfo {year} {2016})}\BibitemShut {NoStop}%
\bibitem [{\citenamefont {Brodin}\ \emph {et~al.}(2000)\citenamefont {Brodin},
  \citenamefont {Marklund},\ and\ \citenamefont {Dunsby}}]{Brodin&al:PRD:2000}%
  \BibitemOpen
  \bibfield  {author} {\bibinfo {author} {\bibfnamefont {Gert}\ \bibnamefont
  {Brodin}}, \bibinfo {author} {\bibfnamefont {Mattias}\ \bibnamefont
  {Marklund}},\ and\ \bibinfo {author} {\bibfnamefont {Peter K.~S.}\
  \bibnamefont {Dunsby}},\ }\bibfield  {title} {\enquote {\bibinfo {title}
  {Nonlinear gravitational wave interactions with plasmas},}\ }\href {\doibase
  10.1103/PhysRevD.62.104008} {\bibfield  {journal} {\bibinfo  {journal}
  {Phys.\ Rev.\ D}\ }\textbf {\bibinfo {volume} {62}},\ \bibinfo {pages}
  {104008} (\bibinfo {year} {2000})}\BibitemShut {NoStop}%
\bibitem [{\citenamefont {Tamburini}\ \emph {et~al.}(2011)\citenamefont
  {Tamburini}, \citenamefont {Thid\'e}, \citenamefont {Molina-Terriza},\ and\
  \citenamefont {Anzolin}}]{Tamburini&al:NPHY:2011}%
  \BibitemOpen
  \bibfield  {author} {\bibinfo {author} {\bibfnamefont {Fabrizio}\
  \bibnamefont {Tamburini}}, \bibinfo {author} {\bibfnamefont {Bo}~\bibnamefont
  {Thid\'e}}, \bibinfo {author} {\bibfnamefont {Gabriel}\ \bibnamefont
  {Molina-Terriza}},\ and\ \bibinfo {author} {\bibfnamefont {Gabriele}\
  \bibnamefont {Anzolin}},\ }\bibfield  {title} {\enquote {\bibinfo {title}
  {Twisting of light around rotating black holes},}\ }\href {\doibase
  10.1038/NPHYS1907} {\bibfield  {journal} {\bibinfo  {journal} {Nature Phys.}\
  }\textbf {\bibinfo {volume} {7}},\ \bibinfo {pages} {195--197} (\bibinfo
  {year} {2011})}\BibitemShut {NoStop}%
\end{thebibliography}

%
\end{document}